\documentclass[journal,twocolumn]{IEEEtran}

\IEEEoverridecommandlockouts                              
\usepackage{cite}
\usepackage{subcaption}
\usepackage{graphicx}
\usepackage{amsmath}
\usepackage{amssymb}
\usepackage[normalem]{ulem}
\usepackage[ruled,vlined]{algorithm2e}
\usepackage{framed}
\usepackage[table]{xcolor}
\usepackage{array}
\usepackage{booktabs, tabularx}
\usepackage{multirow}
\usepackage{bigints}

\usepackage{times}

\usepackage[]{algorithm2e}
\usepackage{bm}
\usepackage{color}
\usepackage{ulem}
\graphicspath{ {images/} }

\newcommand{\prob}{\mathbb{P}}
\usepackage{balance}

\usepackage{float}
\usepackage{caption}

\title{Where to Deploy an Airborne Relay in Unknown Environments: Feasible Locations for Throughput and LoS Enhancement\vspace{0.01cm}}
\author{
\IEEEauthorblockN{Juan David Pabon, Matthew C. Valenti, and Xi Yu
 } \\
West Virginia University, Morgantown, WV, USA. \\
\vspace{-0.25cm}
}

\begin{document}

\maketitle
\thispagestyle{empty}
\pagestyle{empty}


\begin{abstract}
The deployment of heterogeneous teams of both air and ground mobile assets combines the advantages of mobility, sensing capability, and operational duration when performing complex tasks. Air assets in such teams act to relay information between ground assets but must maintain unblocked paths to enable high-capacity communication modes. Obstacles in the operational environment may block the line of sight (LoS) between air assets and ground assets depending on their locations and heights. 
In this paper, we analyze the probability of spanning a two-hop communication between a pair of ground assets deployed in an environment with obstacles at random locations and with random heights (\textit{i.e.} a Poisson Forest) using an air asset at any location near the ground assets. We provide a closed-form expression of the LoS probability based on the 3-dimensional locations of the air asset. We then compute a 3-D manifold of the air asset locations that satisfy a given LoS probability constraint.  We further consider throughput as a measure of communication quality, and use it as an optimization objective.

\end{abstract}

\vspace{-0.2cm}

\section{Introduction}
\label{intro}
\label{introduction}

The use of coordinated autonomous mobile agents, consisting of both ground and air assets, has become prevalent in various applications such as long-term monitoring and post-disaster rescue in large and complex operational environments like urban areas and forests \cite{langelaan2005towards,zheng2010multirobot,harikumar2018multi,couceiro2019semfire,tian2020search,oliveira2021advances}. The deployment of such a heterogeneous set of mobile agents offers several advantages. The air assets provide swift mobility and have large sensing areas, while the ground assets offer high sensing accuracy and long operational duration \cite{grocholsky2006cooperative}. In different scenarios, air assets may act as communication relays \cite{chaimowicz2005deploying} or may provide remote sensing of the battleground \cite{wei2019air}. The locations of these air assets should be carefully planned such that they maintain connectivity with the proper sets of ground assets.

Streaming a large volume of data between assets, especially between air assets and ground assets, is essential for their successful coordination \cite{doriya2015brief}. Modern communication technologies tend to boost the data transmission rate by using very short wavelengths (\textit{e.g.} mmWave frequencies or even visible light), which have compromised penetration capability and are, therefore, easily blocked \cite{hriba2017accurately}. In real-world operational environments, it is common to encounter obstacles of various locations, shapes, and heights that can block the line of sight (LoS) between different assets. Such LoS blockage may also hinder localization and mapping techniques that rely on cameras or lidar \cite{mysorewala2009multi,benjamin2015real}. 

A benefit of using mobile agents is that they can reposition themselves to recover the necessary LoS paths. Assuming that a detailed map of the obstacles is provided, mobile assets can carefully plan their routes through the environment to maintain LoS \cite{gasparetto2015path}. However, such an assumption is usually a luxury for various reasons. Civilian activities, military operations, and disasters may change the landscape of a previously known area. New obstacles can emerge, existing structures may be destroyed or altered, and the terrain can be transformed. Certain environments, like forests, can pose additional difficulties. They may contain a large number of obstacles, such as trees and dense vegetation, which can be challenging to catalog and represent accurately on a map.  Relying solely on pre-existing maps for path planning and obstacle avoidance is imprudent in light of these factors.

An alternative approach to address the limitations of fixed, deterministic maps is to model obstacles in urban areas or forests as distributed randomly.
For instance, the authors of \cite{baccelli2015correlated,Gapeyenko,hriba2021optimization} model the locations of buildings in an urban area using a Manhattan Poisson line process with randomly distributed building heights.   Forested areas were modeled as having randomly located trees in 
\cite{karaman2012high,karaman2012high2,junior2021fast} and with random heights in \cite{kohyama1989frequency,felfili1997diameter,mauro2011influence}. A forest-like cluttered environment can be modeled via a Poisson point process (PPP) and is referred to as a \textit{Poisson forest}.

In environments where the locations of the obstacles and their heights can be modeled as a stochastic process, the probability that the path between two assets crosses an obstacle, thereby blocking the LoS, can be computed accordingly. The so-called \emph{LoS probability} is then the complement of the probability that the path is blocked.  Our previous work \cite{pabon2022air} demonstrated the benefits of deploying an air asset as a communication relay for ground assets by deriving the LoS probability and throughput in a Poisson forest.   However, in that work, we did not solve the problem of where to deploy the air assets, but instead assumed that the air asset was located precisely halfway between the two ground assets.

Deploying an air asset precisely in the middle of a pair of ground assets presents challenges, especially when dealing with moving ground assets or a constantly moving air asset such as a fixed-wing aircraft. Consequently, this paper focuses on the issue of identifying optimal locations for the air asset \emph{away} from the midpoint. Our criterion for finding these locations is based on achieving a constant LoS probability. The nexus of these locations form a curve in 2-D space or a surface in 3-D space, defining a volume within which the LoS probability can be guaranteed. By constraining the air asset's location or path to remain within this volume, the desired performance target is maintained.

In addition to considering locations of constant LoS probability, our study also takes into account \emph{capacity} and \emph{throughput} metrics. More precisely, the capacity is the \emph{Shannon} capacity of the LoS link, while the throughput is defined as the expected value of capacity with respect to the LoS probability. These metrics compensate for distance and discourage infeasible solutions that suggest placing the air asset at very high altitudes.
Similar to LoS probability, the metrics of capacity and throughput can be utilized to define a surface in 3-D space where the values remain constant. As will be shown, the surface of constant throughput is highly nonlinear yet relatively easy to compute using the theory developed in this paper.  Consequently, the contributions presented in this paper significantly simplify the challenging task of planning the locations of air assets. 
The proposed relay placement strategies in this paper offer a moderate level of complexity, which can contribute to improved real-time decision-making and response capabilities.

The paper is organized as follows. Sec.~\ref{section2} formulates the problem. Sec.~\ref{section3} provides an analysis that provides closed-form solutions to determining the positions of an air asset that result in a constant value of LoS probability.   Sec.~\ref{section4} continues the analysis by considering capacity and throughput, similarly finding positions for the air asset that result in constant values of these quantities.   Sec.~\ref{simulation} provides numerical results that illustrate the analysis and allow general trends to be noted.  Finally,  Sec.~\ref{conclusion} concludes the paper.

\section{Problem Formulation}
\label{section2}

Consider a pair of ground assets equipped with communication devices whose heights are equal to $h_g$. These assets are deployed in a operational space ($e.g.$ a forest) with randomly distributed obstacles ($e.g.$ trees).
The location of these obstacles can be described by a two-dimensional PPP with a fixed density $\lambda_f$. This density represents the expected number of \emph{potential} obstacles per unit area\footnote{In this paper, we distinguish between \emph{potential} obstacles and \emph{blockages}.  A potential obstacle is an obstacle of height $H$ that may or may not block the LoS path. On the other hand, a blockage is a potential obstacle that is tall enough to block the path.  Whether the LoS is blocked or not depends on the heights of the obstacle and the communicating devices involved.}. Let $N$ be the number of potential obstacles in an operational environment of area $A_f$, which is referred to as a \textit{Poison forest}. From the properties of a PPP, the probability that the number of potential obstacles in the area is equal to $n$ is
\begin{equation*}
    \mathbb{P} \{N=n\} = \frac{({\lambda_f A_f)}^n}{n!}e^{-\lambda_f A_f}.
\end{equation*}

In a Poison forest, the height of any obstacle is represented by a non-negative random variable $H$. The distribution of $H$ may vary \cite{kohyama1989frequency,felfili1997diameter,mauro2011influence}. Let $F_H(h)$ denote the cumulative distribution function (CDF) of $H$. While the analysis in this paper is not limited to any particular distribution, to illustrate our method and provide closed-form expressions, this paper focuses on the specific realistic cases where $H$ assumes a uniform distribution or a truncated Gaussian distribution. For the uniform distribution, $H \in [0,h_{\max}]$, where $h_{\max}$ is the maximum height that an obstacle could have. The truncated Gaussian is created by taking a Gaussian random variable with mean $\mu$ and standard deviation $\sigma$, and then conditioning the variable on $h\geq 0$.   

\begin{figure}[t]
\centering
\includegraphics[width=0.45\textwidth]{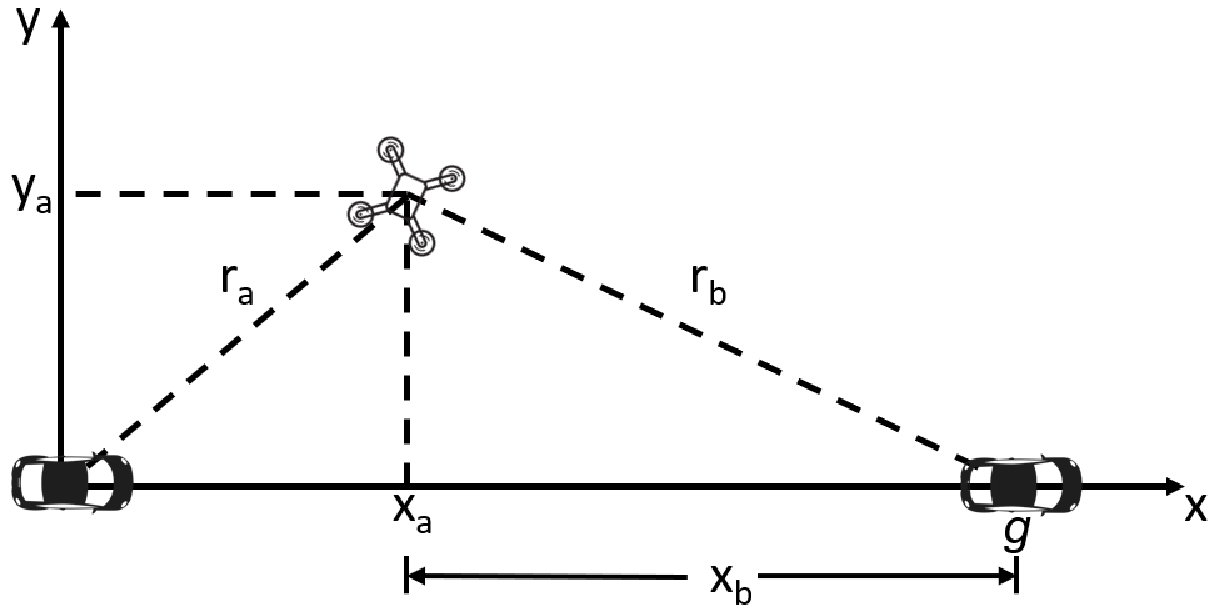}
\caption{{\small Top-view showing the location of the air and ground assets in the ground plane. The ground assets are located at coordinates $(0,0)$ and $(g,0)$, and the air asset is located at an altitude of $h_a$ directly above the coordinate $(x_a,y_a)$. The values $r_a$ and $r_b$ represent the horizontal distances from the ground assets to the air asset.}}
\label{fig:diagram2}
\end{figure}

Let $(x,y)$ denote a location on the ground plane.  The ground assets are assumed to be distance $g$ apart and placed at locations $(0,0)$ and $(g,0)$ on the ground plane as shown in Fig.~\ref{fig:diagram2}. Any potential obstacle with a height above $h_g$ located between the ground assets blocks the LoS between the assets, and hence, becomes a blockage. Consider that the obstacles have a non-trivial thickness $W$ and that the average thickness of the obstacles is $\mathbb{E}(W)$. In the Poisson forest, the distribution of potential obstacles  along the straight line joining any two points can be characterized by a 1-D Poisson process with fixed density $\lambda_0 = \mathbb{E}(W)\lambda_f $ describing the expected number of obstacles per unit length.

When the ground assets are spaced far apart, it is very likely that at least one obstacle will block the LoS of the direct path between them.  To mitigate this issue, an air asset can be deployed as a relay to aid in communication.   Relaying follows a two-hop protocol, with the first hop involving a transmission from the first ground asset to the air asset, and the second hop involving a transmission from the air asset to the second ground asset.  
Let the altitude (or height) of the air asset be $h_a$ and its location, as projected onto the ground plane, be $(x_a,y_a)$ as shown in Fig. 1.  Define the \emph{horizontal distance} between a ground asset and the air asset to be the distance between the ground asset and the projection of the air asset onto the ground plane.   Let $r_a$ be the horizontal distance between the ground asset at $(0,0)$ and the air asset, and $r_b$ be the horizontal distance between the ground asset at $(g,0)$ and the air asset, as shown in Fig. 1.   It follows that $ r_{a} = \sqrt{x_{a}^{2} + y_{a}^2}$ and $r_b = \sqrt{(g - x_a)^2 + y_a^2}$.

In our previous work \cite{pabon2022air}, the LoS probability  is computed considering a fixed position of the air asset (in the middle of the ground assets). Maintaining the air asset at a fixed position can introduce vulnerabilities, as it becomes more susceptible to attacks. Additionally, certain air assets (\emph{e.g.} fixed-wing aircraft or blimps) may lack precise control mechanisms to remain in the same position consistently. Furthermore, the ability to explore a large terrain or facilitate communication among multiple assets can be compromised if the air asset is required to remain in a static position at all times.     

When the air asset has the flexibility to be deployed throughout the entire operational environment, it becomes essential to identify the positions that can deliver the desired performance and fulfill the communication requirements of the ground assets. The performance can be defined in terms of Los probability and throughput, which is a metric that trades off between communication distance and the effect of LoS blockage. In the following sections, we calculate regions in 3-D space identifying where the air relay will provide desired values for LoS probability, capacity, and throughput. Furthermore, the altitude of the air asset is optimized by maximizing the throughput at each possible position in the horizontal plane.

\section{Positions with Constant LoS probability}
\label{section3}
 In \cite{pabon2022air}, the LoS probability for the hop between one of the ground assets, $i \in \{a,b\}$, and the air asset\footnote{Note that the LoS probability is reciprocal; i.e., the LoS probability from a ground asset to an air asset is the same as the LoS probability from the air asset to the same ground asset.} was found to be:
  \begin{equation}
    \prob_{LoS}^{ga}(r_i) = \exp \left(- \int_{0}^{r_{i}} \lambda(r) dr \right).
    \label{eq:Plos}
\end{equation}
$\lambda(r)$ is the density of a PPP that represents blockages; \emph{i.e.}, only those potential obstacles that exceed a \emph{critical height} $h_c$ and therefore may block the LoS path. $h_c$ is the minimum height of an obstacle able to block the ground-air LoS at a given distance $r$ from a ground asset. $h_c$ is distance dependent, increasing with the distance $r$. Because $h_c$ is distance dependent, so is the density $\lambda(r)$, and thus the PPP is inhomogeneous.  

More specifically,  $\lambda(r)$ depends on the density $\lambda_0$ representing the locations of the potential obstacles along with the probability distribution of the obstacles' heights evaluated at the critical height $h_c(\cdot)$. In particular, the density $\lambda(r)$ is determined as follows (see \cite{pabon2022air} for details):
\begin{equation}
\label{eq:InhomoDensity}
    \lambda(r) = \lambda _{0}\left[1-F_{H}\left(h_{c} \left(r \right)\right) \right].
\end{equation}

When the value of $\mathbb{P}^{ga}_{LoS}(r_i)$ is fixed, $i.e.$ equal to a constant within the interval $(0,1]$, \eqref{eq:Plos} can be solved for $r_i$. The solution provides the horizontal distance between the ground and the air asset that results in the fixed value of $\mathbb{P}^{ga}_{LoS}(r_i)$. Once solved, the locations of constant LoS probability for a single hop form a circle around the ground asset. This circle has a radius of $r_i$, and by restricting the movement of the air asset to fly along this circular path, a constant LoS probability can be maintained.

However, in this paper, we are concerned with using the air asset to provide a relay connecting two ground assets.  For the two ground assets to be connected, we assume that there must be a clear LoS from the air asset to \emph{both} of the ground assets.  Define 
$\prob_{LoS}^{gag}(r_a, r_b)$ to be the probability of LoS for two-hop ($i.e.$, ground-air-ground) communication, which requires a LoS from the first ground asset to the air asset and a LoS from the air asset to the second ground asset.   Assuming that the two paths are independent (a reasonable assumption unless their angular distance is close, see \cite{hriba2018impact}), this probability is the product of the LoS probabilities of each of the two hops, $i.e.$,  
\begin{equation}
   \prob_{LoS}^{gag}(r_a, r_b) = \prob_{LoS}^{ga}(r_a)\prob_{LoS}^{ga}(r_b).
    \label{eq:Plos2}
\end{equation}

As with the case for one ground asset, it is possible to fix the value of LoS probability in \eqref{eq:Plos2} and find the location of the air asset that provides a constant value. However, because there are now two ground assets, the region of constant LoS probability will become an ellipse rather than a circle, as will be shown subsequently for height distributions that are uniform or truncated Gaussian.

\subsection{LoS for Uniform Distribution}

When $H$ is uniform over $[0,h_{\max}]$, the end-to-end LoS probability can be expressed as
\begin{equation}
 \prob_{LoS}^{gag}(r_a, r_b)=e^{-\Lambda_{u}(r_a)}e^{-\Lambda_{u}(r_b)}
 \label{eq:Plos_ab}
\end{equation}
where $\Lambda_{u}(\cdot)$ is the result of the integral in \eqref{eq:Plos}. Assuming that the air asset is deployed at $h_a > h_{\max}$,  $\Lambda_{u}(\cdot)$ is equal to

\begin{equation*}
\Lambda_{u}(r_i) =\lambda_{0}r_{ci}\left[ 1 - \dfrac{h_{g}}{h_{\max}} - \left( \dfrac{h_{a} -h_{g}}{2r_{i}h_{\max}}\right)r_{ci} \right],
\end{equation*}
where $r_{ci}$ for $i \in \{a,b\}$ is the critical distance beyond which the critical height is taller than $h_{\max}$.  It is found to be:
\begin{equation*}
    r_{ci} = r_{i}\left(\dfrac{h_{\max} - h_g}{h_a - h_g}\right).
\end{equation*}

Fixing the value of $\prob_{LoS}^{gag}(r_a,r_b)=P$ and rewriting \eqref{eq:Plos_ab}, the following equation is obtained:
\begin{equation}
     r_a + r_b= c_u,
    \label{eq:distances_uniform}
\end{equation} 
where $c_u$ is 

found to be:
\begin{equation*}
    c_u = \dfrac{-(h_a - h_g)\ln(P)}{\lambda_0(h_{\max} - h_g)\left( 1 - \dfrac{h_{g}}{h_{\max}} - \dfrac{h_{\max} -h_{g}}{2h_{\max}}\right)}.
\end{equation*}

Using \eqref{eq:distances_uniform}, the relation between the distance from the air asset to each ground asset and the desired value of $P$ can be obtained. For a fixed $h_a$, \eqref{eq:distances_uniform} describes an ellipse since the sum of the distances from the air asset to the ground assets $a$ and $b$ is constant, \emph{i.e.}, the value $c_u$.

In \eqref{eq:distances_uniform}, $r_a$ and $r_b$ can be expressed in terms of the position of the air asset in the ground plane. Then, the relation between the location of the air asset and the desired value of $P$ can be obtained. In Fig.~\ref{fig:diagram2}, it is observed that the coordinate $(x_a,y_a)$ is the location of the air asset in the ground plane.  By replacing $(r_a,r_b)$ with the corresponding values of $(x_a,y_a)$, \eqref{eq:distances_uniform} can be written as

\begin{equation}
     \sqrt{x_a^2 + y_a^2} + \sqrt{(g-x_a)^2 + y_a^2}= c_u.
    \label{ellipse1}
\end{equation} 

After simplifying \eqref{ellipse1} and substituting $(x,y) = (x_a,x_b)$, the following equation is obtained
\begin{equation}
    \dfrac{4}{c_u^2}\left( x -\dfrac{g}{2}\right)^2 + \dfrac{4}{c_u^2-g^2}y^2 = 1.
    \label{ellipse_uniform}
\end{equation}
When the air asset flies above coordinate $(x,y)$, it will have the desired two-hop LoS probability $P$. When $(c_u^2 - g^2) >0$, the locations of $(x,y)$ allowed by \eqref{ellipse_uniform} form an ellipse with a major axis along the x-axis and foci located at $(0,0)$ and $(g,0)$; \emph{i.e.,} at the positions of the ground assets.

\subsection{LoS for Truncated Gaussian Distribution}
Now consider the case that $H$ is a truncated Gaussian. To find the positions of the airborne relay producing a constant two-hop LoS probability, start with \eqref{eq:Plos}. For the truncated Gaussian distribution, the integral of \eqref{eq:Plos} is equal to $ r_i\left(c + c k\right)$, where {\small$c=\lambda_0/(2\Phi(\mu/\sigma))$} and $\Phi(\cdot)$ is the CDF of the standard normal distribution. The value of $k$ is given by
\begin{equation*}
    k = \left( \dfrac{ e^{-a^2} + \sqrt{\pi}\left[(d-a)\text{erf}(a-d) + a\text{erf}(a) \right] - e^{-(a-d)^2}}{\sqrt{\pi}d} \right)
\end{equation*}
where erf($\cdot$) is the error function, {\small$a=(\mu - h_g)/(\sqrt{2}\sigma$) and $d = (h_a-h_g)/(\sqrt{2}\sigma)$}. Substituting the result of this integration into \eqref{eq:Plos}, the following equation is obtained: 
\begin{equation*}
    \mathbb{P}_{LoS}^{ga}(r_i) = e^{-r_i(c+ck)}.
\end{equation*}

It follows from \eqref{eq:Plos2} that the two-hop LoS probability is equal to
\begin{equation*}
    \mathbb{P}_{LoS}^{gag}(r_a,r_b) = e^{-(r_a + r_b)(c+ck)}.
\end{equation*}
Fixing the desired value of $\mathbb{P}_{LoS}^{gag}=P$ gives
\begin{equation*}
    P = e^{-(r_a + r_b)(c+ck)}.
\end{equation*}
Rewriting this equation gives
\begin{equation}
    r_a + r_b = c_t
    \label{distances_truncated_g}
\end{equation}
where $c_t = -\ln(P)/(c+ck)$. 
As with \eqref{eq:distances_uniform}, equation \eqref{distances_truncated_g} describes an ellipse,
only now the two distances sum to a different value; \emph{i.e.,} they sum to $c_t$.

By expressing $r_a$ and $r_b$ in terms of the position of the air asset in the ground plane $(x,y) = (x_a, y_a)$, and assuming $g<c_t$, simplifying \eqref{distances_truncated_g} yields
 \begin{equation}
    \dfrac{4}{c_t^2}\left( x -\dfrac{g}{2}\right)^2 + \dfrac{4}{c_t^2-g^2}y^2 = 1.
    \label{ellipse_truncated_g}
\end{equation}
As with the uniform distribution, air assets flying above coordinate $(x,y)$ will provide the desired two-hop LoS probability $P$, and for fixed $h_a$, the locus of all $(x,y)$ forms an ellipse.

\section{Link Capacity and Expected Throughput}
\label{section4}
Sec. \ref{section3} found the positions of the air asset that produce a fixed two-hop LoS probability. However, the LoS probability does not consider the signal power loss that occurs as the distance between the air asset and the ground asset increases. 
To provide a more comprehensive assessment of communication performance, this section introduces the computation of capacity and throughput metrics. These metrics take into account the distance-dependent signal power loss, allowing for a more suitable objective when determining the optimal placement of the airborne relay.

\subsection{Positions with Constant Link Capacity}

As a measure of capacity, we specifically consider the \emph{Shannon} capacity. The Shannon capacity represents the maximum achievable data rate for an unblocked link. It quantifies the upper limit of information transmission rate over the channel, taking into account such factors as the channel propagation model, distance transmitted, transmitted power, transmission bandwidth, and antenna gains.  By considering the Shannon capacity, we can assess the maximum data rate that can be achieved under ideal conditions for a given communication link, such as between the air asset and a ground asset.

The (Shannon) capacity of the link between ground asset $i \in \{a,b\}$ and the air asset can be written as
\begin{equation}
    C_i = B\log_2(1+\mathsf{SNR}),
    \label{capcity}
\end{equation}
where $B$ is the signal bandwidth and $\mathsf{SNR}$ is the signal-to-noise ratio.  When $B$ is in units of Hertz, $C_i$ is in units of bits-per-second (bps).  When expressed in dB, the value of $\mathsf{SNR}$ is given by
\begin{equation}
    \mathsf{SNR}^{\text{dB}} =\mathsf{SNR}^{\text{dB}}_0 - 10\alpha\log_{10}\left(\dfrac{d_i}{d_0}\right),
\end{equation}
where $d_i$ is the Euclidean distance from the air asset to ground asset $i$, $\alpha$ is the path-loss exponent, and $d_0$ is the reference distance. $\mathsf{SNR}^{\text{dB}}_0$ is the signal-to-noise ratio when the transmission distance is equal to $d_0$ assuming free-space propagation up to that distance. Note that, while it depends on the communication distance, the capacity does not depend on the LoS probability.

For the ground asset located at the origin ($i=a$), the Euclidean distance $d_a^2 = x^2+y^2+h_a^2$, where $(x,y,h_a)$ is the 3-D position of the air asset.  Rearranging
\eqref{capcity} yields:
\begin{equation}
    x^2+y^2+h_a^2 = d^2_0\left(\dfrac{10^{\mathsf{SNR}_0^{\text{dB}}/10}}{2^{C_{i}/B}-1} \right)^{2/\alpha}.
    \label{eq:shape_Capacity}
\end{equation}

For the other ground asset ($i=b$), which is located at coordinate $(g,0)$, the term $x$ in (\ref{eq:shape_Capacity}) must be replaced by $g-x$ per Fig. 1.

As with the LoS probability,  \eqref{eq:shape_Capacity} shows that the airborne relay should be located on a circle around the ground asset when there is just a single hop.   However, when two-hop (ground-air-ground) relaying is considered, the end-to-end capacity is limited by the minimum of the capacities of the two hops.  Moreover the duplexing operation at the relay must be taken into account. Thus, the two-hop capacity is $C = \min(C_a,C_b)/2$, where the multiplication by $1/2$ accounts for the time-division duplexing operation at the relay. 

It follows that capacity is determined by the shorter of the two links.  Hence, when the air asset is closer to the ground asset located at the origin, $d_a < d_b$ and $C_a > C_b$.   Conversely, when $d_b < d_a$, then $C_b > C_a$. Since $d_a < d_b$ when $x < g/2$, the capacity for the ground-air-ground link is given by 

\begin{equation}
    C = \frac{1}{2}\min(C_a,C_b) = 
     \begin{cases}
        C_b/2 & \text{if } x \leq g/2\\
        C_a/2 & \text{if } x > g/2.  
    \end{cases}
    \label{eq:capacity}
\end{equation}

\subsection{Positions with Constant Throughput}
The LoS probability serves as a valuable indicator for predicting the presence or absence of a link between the assets, while the capacity provides an estimation of the maximum achievable data rate if a link exists. However, neither of these parameters alone provides a comprehensive assessment of the overall communication quality. The LoS probability accounts for blockages but does not consider the impact of transmission distance on signal power loss. On the other hand, capacity considers the transmission distance but overlooks the blockage process.   An appropriate metric for assessing the quality of the link is the \emph{throughput}, which strikes a balance between LoS probability and capacity \cite{pabon2022air}. 

The capacity provided in \eqref{capcity} is the capacity of a given link when it is unblocked.  When the link is blocked, its capacity is zero.   Thus, in environments with random blocking, the capacity of link $i$ is a random variable that assumes a value of $C_i$ with probability $\mathbb{P}_{LoS}^{ga}$ and a value of zero with  probability $1-\mathbb{P}_{LoS}^{ga}$.   The throughput of a link in the presence of blocking is the expected value of the capacity of the link, where the expectation is with respect to the LoS probability.  In particular, the throughput of a single hop is
\begin{eqnarray}
    T_i 
    & = &
    \mathbb{P}_{LoS}^{ga}C_i.
\end{eqnarray}

For a two-hop communication, the throughput is the expectation of the end-to-end capacity, where the expectation is with respect to the two-hop LoS probability; \emph{i.e.,} it is as follows:
\begin{equation}
    T =
    \begin{cases}
        \mathbb{P}_{LoS}^{gag}C_b/2 & \text{if } x  \leq g/2.\\
         \mathbb{P}_{LoS}^{gag}C_a/2 & \text{if } x > g/2.
    \end{cases} 
    \label{eq:throughput}
\end{equation}

Fixing the value of throughput in \eqref{eq:throughput} and allowing the 3-D position $(x,y,h_a)$ of the air asset to vary, we can find a surface in 3-D space that guarantees the desired throughput. Also, the positions of the air asset that maximize the throughput can be found, for instance, by fixing the coordinate on the ground plane $(x,y)$ and determining the altitude $h_a$ that maximizes the throughput.

\section{Numerical Results}
\label{simulation}
This section presents numerical results to illustrate the theoretical concepts discussed in Sections \ref{section3} and \ref{section4}. The target values of LoS probability, capacity, and throughput are varied to determine the 2-D and 3-D surfaces for which these values are constant.  This provides insight into possible flight paths for the air asset that provide the necessary performance metrics.   

Unless otherwise specified, it is assumed that the values of the key physical parameters are assumed to be $\lambda_0 = 0.02$, $h_g=2$~m,  $B=100$~MHz, $d_0=1$~m, $\mathsf{SNR}_0^{\text{dB}}=50$~dB, and $\alpha=2.3$.  This path-loss coefficient corresponds to measured LoS pathloss at $38$ GHz \cite{Rappaport2013}.  The  value of $\mathsf{SNR}_0^{\text{dB}}$ corresponds to a carrier frequency of 38 GHz, a bandwidth of 100 MHz, a transmit power of $0$ dBm, a receiver noise figure of $11$ dB, and antenna gains of $12.1$ dBi for both the transmit and receive antennas, which are the gains reported for a compact 6-element array operating at $38$ GHz in \cite{Rahayu2018}.   Both kinds of obstacle height distributions are considered; for the case that the heights are uniform we use 
$h_{\max}=29$~m and for the case that they are truncated Gaussian we use $\mu=19$~m and $\sigma=10$~m.

\subsection{LoS Probability}

This section shows the results obtained when the desired LoS probability $P$ and the distance between ground assets $g$ take different values in \eqref{ellipse_uniform} and \eqref{ellipse_truncated_g}. 

\begin{figure}[t]
\centering
\includegraphics[width=0.52\textwidth]{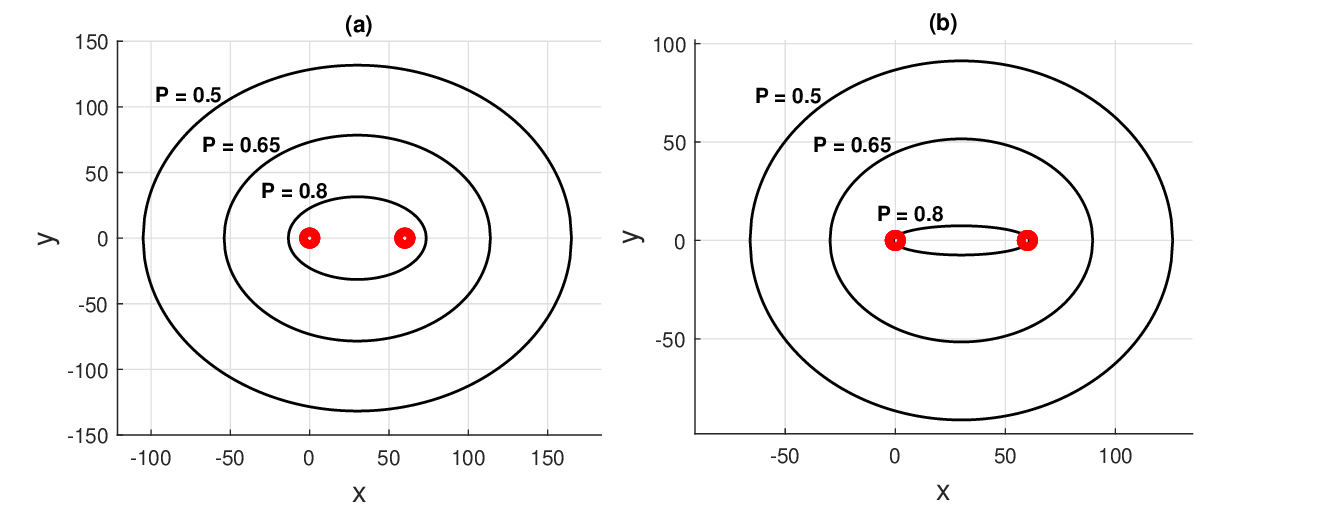}
\caption{{\small Air asset positions producing fixed LoS probability. Obstacle heights are (a) uniform distribution and (b) truncated Gaussian distribution. Red dots indicate the location of the ground assets.}}
\label{fig:LoS_paths}
\end{figure}

Fig.~\ref{fig:LoS_paths} considers the case that $h_a = 100$~m and $g=60$~m.  In this figure, red dots indicate the location of the two ground assets, while the black ellipses show the locations for the air asset that provide constant LoS probabilities equal to $P=0.8$ (inner ellipse), $P=0.65$ (middle ellipse), and $P=0.5$ (outer ellipse).   The subplot on the left corresponds to the case that the obstacle height distribution is uniform while the subplot on the right corresponds to the case that it is truncated Gaussian. For greater LoS probability the eccentricity of the ellipses increases and for smaller probabilities the eccentricity decreases and the major axis of the ellipse increases its length.

\begin{figure}[t]
\centering
\includegraphics[width=0.45\textwidth]{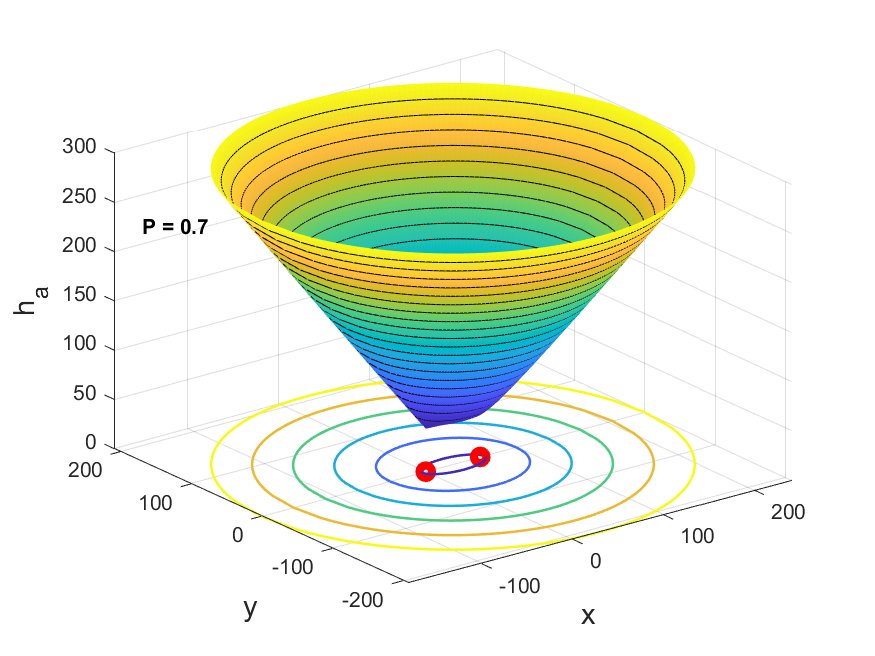}

\caption{{\small Air asset positions producing $P=0.7$. The contour plot of the surface is shown in the x-y plane. Red dots indicate the location of the ground assets. Uniform height distribution is considered.}}
\label{fig:surface_uniform}
\end{figure}

When $h_a$ changes and the same LoS probability is required, the elliptic cone presented in Fig.~\ref{fig:surface_uniform} is obtained. This surface represents the positions that produce the desired LoS probability $P$. In this case, $P=0.7$ and the uniform obstacle height distribution is considered. The minimum altitude $h_a$ at which the air asset can be deployed to  produce the desired $P$ with a given distance $g$ between ground assets can be determined by equating $c_u$ to $g$ and solving for $h_a$. This minimum altitude determines the distance from the bottom of the elliptic cone to ground. At the bottom of the cone $h_a = 44.29$~m.  

\begin{figure}[t]
\centering
\includegraphics[width=0.37\textwidth]{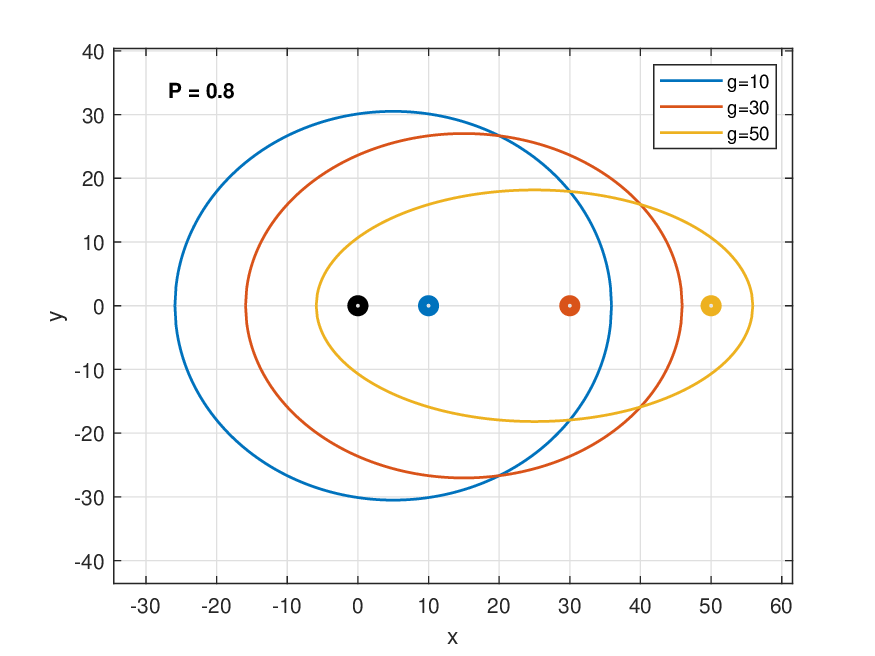}
\caption{{\small Ellipses representing air asset positions producing $P=0.8$. The black dot represents the position of ground asset $a$. Dots with different colors represent positions of ground asset $b$ and each dot has an ellipse with the same color.  Truncated Gaussian height distribution is considered.}}
\vspace{-0.6cm}
\label{fig:gaussian_2}
\end{figure}

Fig.~\ref{fig:gaussian_2} shows the positions of the air asset that guarantee a fixed LoS probability for three different values of distance between ground assets ($g=\{10,30,50\}$ m) when $h_a=100$~m. In this case, the obstacle height distribution is truncated Gaussian.  When the distance between the ground assets is small, the ellipse has little eccentricity (i.e., it is almost a circle), while the eccentricity grows with increasing $g$.   As $g$ tends to $c_t$ (or, for uniform height distributions, $h_u$), the positions for the air asset will tend towards a straight line connecting the ground assets. That distance $g$ will be the maximum distance that the assets could separate while the air asset is able to guarantee the desired LoS probability.

\subsection{Capacity and Throughput}
\begin{figure}[t]
\centering
\includegraphics[width=0.5\textwidth]{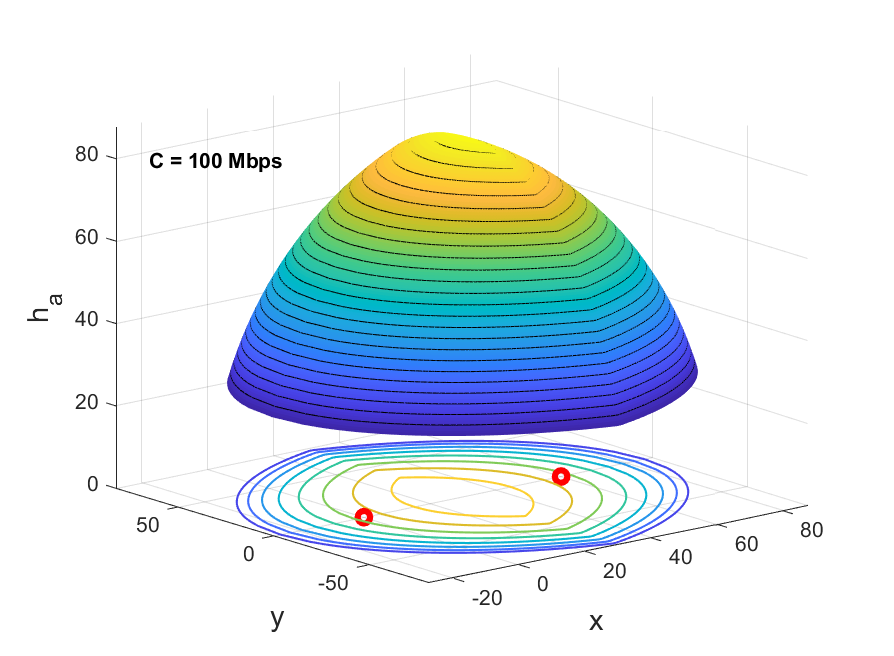}
\caption{{\small Air asset's positions producing $C=100$~Mbps. A contour plot is projected onto the xy-plane.}}
\label{fig:capacity}
\end{figure}

In Fig.~\ref{fig:capacity}, the value of capacity in \eqref{eq:capacity} is fixed at $C = 100$ Mbps, and the 3-D region of constant capacity is shown.   When the air asset's position is such that $x\leq g/2$, the capacity of the ground-air-ground link is limited by the capacity of the link between the air asset and ground asset $b$, and the surface is obtained evaluating \eqref{eq:shape_Capacity} at $r_b$. Similarly, for $x>g/2$, the capacity is determined by the link between ground asset $a$ and the air asset, and the surface is obtained evaluating \eqref{eq:shape_Capacity} at $r_a$. Any position inside the volume covered by this surface will produce a capacity greater than 100~Mbps.   Because capacity does not account for the presence of obstacles, the region does not depend on the obstacle height distribution.     

\begin{figure}[t]
\centering
\includegraphics[width=0.5\textwidth]{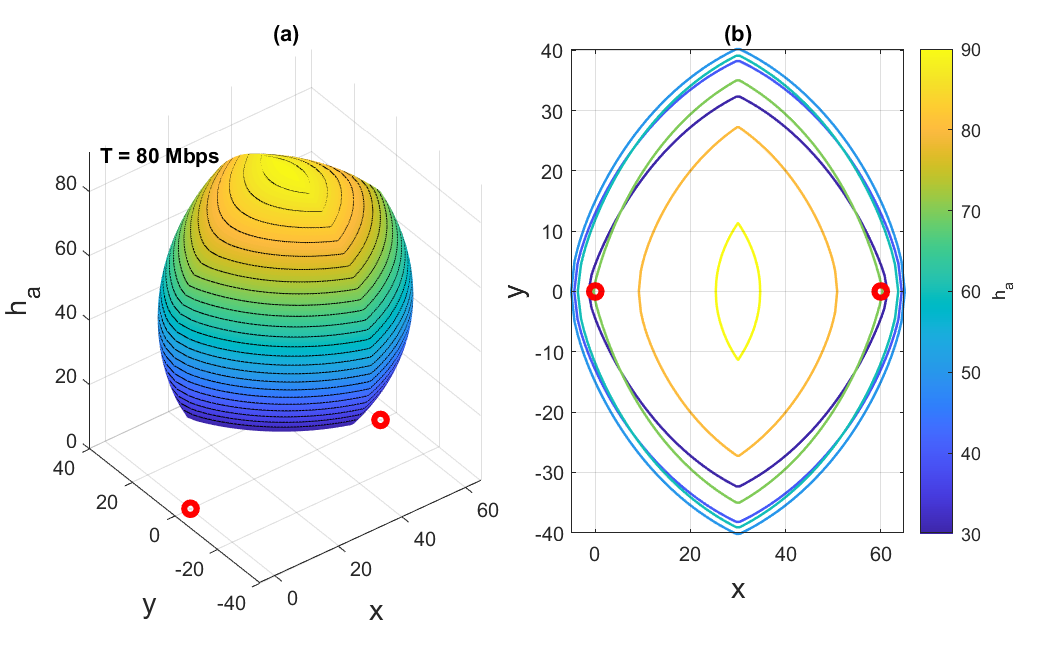}
\caption{{\small (a) air asset's positions producing $T=80$~Mbps. (b) contour plot of the surface shown in (a). The obstacle heights are uniformly distributed.}}
\label{fig:Throughput}
\end{figure}

Next, we consider regions of constant throughput, as it is a metric that balances capacity with LoS probability. Fig.~\ref{fig:Throughput} shows a 3-D surface representing the positions of the air asset that guarantee a throughput equal to 80~Mbps when the uniform height distribution is considered.  The following observations can be made about the surface that is shown. As shown in \ref{fig:surface_uniform}, it can be observed that as the altitude $h_a$ of the air asset increases, the contours of constant LoS probability expand, indicating a larger area covered by LoS connections. On the other hand, in Figure~\ref{fig:capacity}, as $h_a$ increases, the contours of constant capacity shrink. Since the throughput is the product of LoS probability and capacity, in Fig.~\ref{fig:Throughput}, both behaviors are observed, with the cross section areas initially increasing with $h_a$ as the regions of constant LoS probability expand.  But then, after a certain value of  $h_a\geq 53.6$~m, the area of the cross sections decreases as the constant-capacity contours contract with increasing $h_a$. The volume of the region contained by the surface is inversely proportional to the throughput; i.e., if a smaller throughput were considered, then the region shown would be larger.

\begin{figure}[t]
\centering
\includegraphics[width=0.5\textwidth]{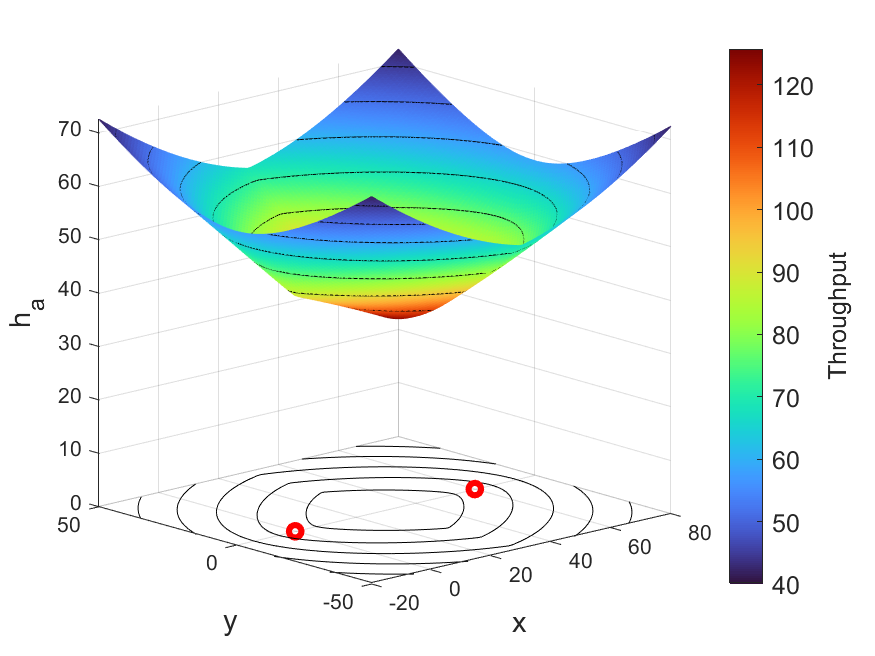}
\caption{{\small Air asset positions producing the maximum throughput. The color of the surface represents the value of throughput indicated by the color bar. The contours in the xy-plane represent the contours for the same air asset altitude. The obstacle heights are uniformly distributed.}}
\label{fig:max_throughput}
\end{figure}

In addition to identifying regions of constant throughput, it is also possible to optimize equation \eqref{eq:throughput} with respect to the altitude $h_a$ of the air asset. This optimization process determines the air asset altitude that maximizes the throughput for a given position in the ground plane. Fig.~\ref{fig:max_throughput} shows the air asset altitude that maximizes throughput for each position of the air asset over the ground plane. It is observed that the maximum possible throughput is obtained when the air asset is located in $(g/2, 0, 36.3)$ for $g=60$~m. The surface shown in Fig.~\ref{fig:max_throughput} allows us to determine the positions across which the air asset should move if it is required to obtain the maximum possible throughput for any of the positions. Additionally, the contours for different heights of the surface in Fig.~\ref{fig:max_throughput} are shown in the xy-plane. 
	
\section{Conclusions and future work}
\label{conclusion}
In this paper, we have developed a framework for determining feasible locations to deploy an airborne relay to enhance communication between two ground assets while satisfying performance metrics including LoS probability, capacity, and throughput. As a metric, throughput strikes a balance between deploying the air asset at locations that increases LoS probability while simultaneously decreases signal power loss due to distance.

A key contribution of this paper is an approach for calculating the throughput based on the location of the air asset. This methodology takes into account the heights of obstacles and their locations described by an inhomogeneous PPP. Closed-form expressions are derived for determining the regions where the air asset can be deployed to achieve desired values of LoS probability or capacity. Knowledge of the maximum achievable throughput facilitates the deployment of the air asset in locations that optimize communication performance, and can be used to identify the optimal altitude to deploy the air asset.

Future research directions include investigating optimal trajectory planning for the air asset when the ground assets are in motion. Additionally, optimizing trajectory planning to deploy the air asset in scenarios involving multiple ground assets requiring communication would be a valuable area of exploration.

\bibliographystyle{IEEEtran}
\balance
\bibliography{sections/library.bib}

\end{document}